\pdfoutput=1

\documentclass[conference]{IEEEtran}
\ifCLASSINFOpdf
  \usepackage[pdftex]{graphicx}
\else
\fi
\usepackage{xcolor}
\usepackage[normalem]{ulem}

\begin{document}

\twocolumn[
\begin{@twocolumnfalse}

\copyright ~2021 IEEE. Personal use of this material is permitted. Permission from IEEE must be obtained for all other uses, in any current or future media, including reprinting/republishing this material for advertising or promotional purposes, creating new collective works, for resale or redistribution to servers or lists, or reuse of any copyrighted component of this work in other works.\\ 

\end{@twocolumnfalse}
]

%
\title{DeepScaleTool : A Tool for the Accurate Estimation of Technology Scaling in the Deep-Submicron Era}


\author{\IEEEauthorblockN{Satyabrata Sarangi and Bevan Baas}
\IEEEauthorblockA{Department of Electrical and Computer Engineering\\
University of California, Davis\\
Email: \{ssarangi, bbaas\}@ucdavis.edu}
}


%


\maketitle

\begin{abstract}

The estimation of classical CMOS ``constant-field'' or ``Dennard'' scaling methods that define 
scaling factors for various dimensional and electrical parameters have become less accurate in the deep-submicron regime, which drives the need for better estimation approaches especially in the educational and research domains. 
We present \emph{DeepScaleTool,} a tool for the accurate estimation of deep-submicron technology scaling by modeling and curve fitting published data by a leading commercial fabrication company for silicon fabrication technology generations from 130~nm to 7~nm for the key parameters of area, delay, and energy.  
Compared to 10~nm--7~nm scaling data published by a leading foundry, the DeepScaleTool achieves an error of 1.7\% in area, 2.5\% in delay, and 5\% in power. 
This compares favorably with another leading academic estimation method that achieves an error of 24\% in area, 9.1\% in delay, and 24.9\% in power.

\end{abstract}


%
\IEEEpeerreviewmaketitle

\section{Introduction}


 

Moore's law~\cite{Moore} has been pivotal in the advancement of the semiconductor industry for decades, which lays out a projection of doubling of the transistors on an IC every two years. Similarly, Dennard scaling~\cite{Dennard, Dennard2} pioneered the progress by showing scaling across physical dimensions, substrate doping, and supply voltage, which in turn results in lower area, delay, and power dissipation for MOSFETs. The resulting changes due to scaling are depicted in terms of an entity called scaling factor~\ensuremath{K}, which is defined as the ratio of 
two technology nodes. These scaling factors are discussed in various textbooks~\cite{rabaey2003digital,Uyemura} and the literature~\cite{Bohr, Stillmaker201774}, and are shown in Table~\ref{tab:scalingdata}.  The key points from the traditional scaling factors shown in Table~\ref{tab:scalingdata} are the following: transistor physical dimensions shrink down by the scaling factor $K$, which in turn scales down the transistor area by a factor of~{\ensuremath{K^2}}. Similarly, the speed of the transistor increases by a factor of~\ensuremath{K} as the delay 
reduces by 
{\ensuremath{1{/}K}}. 

The traditional scaling factors were accurate until the advent of the deep-submicron era. As the transistors get smaller in the deep-submicron regime, due to short channel effects, effect of leakage current and thermal runaway, and process variation, 
the traditional scaling estimations are no longer accurate~\cite{Stillmaker201774, Kuhn}. Moreover, resulting performance gain over recent technology generations is minimal unlike the predictions by traditional scaling estimations.

The inaccuracy in the traditional scaling factors can also be depicted from the real silicon data from various foundries across technology fabrication generations. Notably, Holt~\cite{Holt} discusses 
transistor scaling and its effect on transistor area, gate delay, switching energy, and energy delay product in the deep-submicron regime based on Intel's data. Bohr and Young~\cite{Bohr} 
describe 
Intel's scaling trends for area, transistor performance, 
and
cost per transistor over the past decade. A method to propose accurate scaling predictions has been demonstrated using data from  PTM~\cite{PTM}, ITRS~\cite{ITRS}, and simulated measurements of Fan Out 4, or FO4 circuit~\cite{Stillmaker201774}. However, accuracy achieved in this method differs significantly from both TSMC~\cite{Cai} based silicon technology scaling 
data  
and estimations presented in this article, which is discussed in Section IV.  

\begin{table}[tb]
	\caption{MOSFET device parameters and traditional scaling factors}
	\label{tab:scalingdata}
	\vskip 0.15in
	\begin{center}
		\begin{small}
			\begin{sc}
				\begin{tabular}{l|l}
					\hline
					 \textbf{Parameter} & \textbf{Scaling factor}\\

\hline
Device dimension ~{\ensuremath{(W, L, tox)}} & ~{\ensuremath{1{/}K}}\\
\hline
Doping concentration ~{\ensuremath{Na}} & ~{\ensuremath{K}}\\
\hline
Voltage ~{\ensuremath{V}} & ~{\ensuremath{1{/}K}}\\
\hline 
Current ~{\ensuremath{I}} & ~{\ensuremath{1{/}K}}\\
\hline
Capacitance~{\ensuremath{\epsilon{A{/}t}}} & ~{\ensuremath{1{/}K}}\\
\hline
Delay time {\ensuremath{VC{/}I}} & ~{\ensuremath{1{/}K}}\\
\hline
Power dissipation ~{\ensuremath{VI}} & ~{\ensuremath{1{/}{K^2}}}\\
\hline
Power density ~{\ensuremath{VI{/}A}} & ~{\ensuremath{1}}\\
\hline
\end{tabular}
\end{sc}
\end{small}
\end{center}
\vskip -0.1in
\end{table}

The prediction of accurate scaling factors is also important for a fair comparison of design performance and other metrics across different technology fabrication nodes. Although the ITRS and PTM based simulation and modeling approach looks viable for predicting scaling factors in the deep-submicron regime, supply voltage information is not always publicly released, which is accounted for modeling the delay and power scaling equations in the article~\cite{Stillmaker201774}. Moreover, such estimation method doesn't necessarily align with the actual silicon technology scaling trends, which 
are 
discussed in Section IV. 

In academia,
popular  
textbooks~\cite{rabaey2003digital,Uyemura} that 
cover  
digital VLSI design and scaling of CMOS transistors describe traditional scaling factors and reasons associated with the discontinuity in traditional scaling trends. However, an accurate estimation of scaling factors across technology fabrication nodes and correlation between traditional scaling factors and scaling factors resulting from actual silicon are usually not covered. Therefore, we propose a scaling tool whose modeling is based on the industrial technology scaling trends and polynomial based curve fitting approach for an easy and accurate estimation of scaling factors in the deep-submicron era. 

The major contributions of our work are as follows:
\begin{itemize}
    \item We demonstrate DeepScaleTool, a 
    spreadsheet-based  
    tool for accurate estimation of scaling factors for area, delay, and energy from 130~nm to 7~nm for educational and research purposes.
    \item We show and analyze the percentage errors in between classical and estimated scaling factors from real silicon data in the deep-submicron range.
    \item We also illustrate examples of scaling factors estimation using DeepScaleTool and compare the results both with PTM and ITRS based modeling~\cite{Stillmaker201774} and scaling data from TSMC~\cite{Cai}. 
\end{itemize}

The remainder of the paper is organized as follows. Section~II describes the published silicon technology scaling trends~\cite{Bohr, Holt}, method of scaling data modeling adopted in this work, and the tool framework. Section~III describes the usage of DeepScaleTool and examples of scaling factors computations using the tool. Section~IV discusses comparison of traditional scaling factors vs. accurately estimated scaling factors and differences in scaling estimation methods. Section~V concludes the paper. 

\section{Transistor Scaling Trends, Data Modeling, and DeepScaleTool Framework}

{\begin{figure}
{\centering{\includegraphics[scale = 0.38]{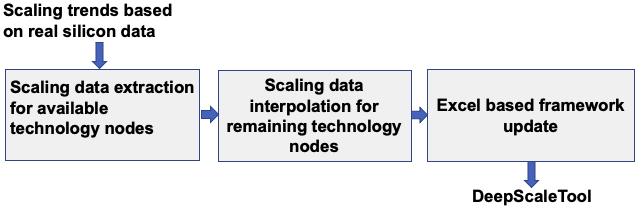}}}
\caption{Procedure of Data Collection and Framework}
\label{fig:procedure}
\end{figure}}
Figure~\ref{fig:procedure} provides an overview of the steps leading to the design of DeepScaleTool. We analyze published transistor scaling trends~\cite{Bohr, Holt}, curve fit scaling data that are available for certain technology nodes using second-order polynomial based models, and then extrapolate scaling data for the rest of the technology nodes. Finally, we design and update the 
spreadsheet-based
framework using modeled scaling factors for a combination of starting and target technology nodes.

\subsection{Transistor Scaling Trends}
The following two notable transistor scaling trends that are based on Intel's silicon results have been analyzed for our work. Holt~\cite{Holt} discusses generational technology benefits over reduction in gate delay, switching energy, and energy delay product. All the presented scaling data in the article span around 65~nm to 10~nm technology nodes and are relative to 65~nm. Moreover, the normalized transistor area across 130~nm to 14~nm technology nodes have been demonstrated. The scaling trends shown in the article over technology fabrication generations infer the following for circuits---acceleration in transistor density, higher performance, and lower power.  

Similarly, Bohr and Young~\cite{Bohr} discuss 
scaling logic circuit area from 45~nm to 10~nm. The key takeaway from the presented circuit area scaling data is that with the advent of newer technology nodes and transistor level innovations more aggressive scaling is possible than the traditional scaling estimation. For example, both 14~nm and 10~nm technology nodes achieve 0.37 times logic area scaling than the previous generation. This article also presents the trends in improved transistor performance, active power, and performance per watt metrics. However, due to unavailability of proper axis labeling the corresponding trends have not been considered for data modeling purposes in this work. 

\subsection{Data Extraction and Modeling}

The g3data~\cite{g3data} tool has been used to extract the digitized data from the plots with technology scaling trends shown in the articles~\cite{Bohr, Holt}. The plots with proper axis labeling have been considered for data extraction. To obtain scaling trends from 130~nm to 7~nm for key circuit parameters like area, delay, and energy, available scaling data across technology fabrication generations have been extrapolated to obtain the scaling data of the corresponding parameter at the missing technology generations. The polynomial based extrapolation models that are used to curve fit for various circuit parameters yield a coefficient of determination or~{\ensuremath{R^2}} value of equal or greater than 0.99. The small differences in area scaling factors that are obtained after modeling the scaling trends in both articles~\cite{Bohr, Holt}, have been offset by taking the average of the corresponding two scaling factors for any given starting and target technology generations.  

\subsection{DeepScaleTool Framework}

Instead of providing big tables consisting of scaling factors across technology fabrication nodes for each circuit parameter, we present a 
spreadsheet-based  
framework for the automated generation of scaling factors for various circuit parameters. The framework is designed using visual basic for applications (VBA) programming language. The scaling factor values fields in the spreadsheet are programmed for any of the supported current and target technology nodes.
The DeepScaleTool is available as an open source tool and it can be accessed at https://sourceforge.net/projects/deepscaletool/~\cite{DeepScale}.
Figure~\ref{fig:tool} depicts a screenshot of the tool. The tool can be updated easily for future nodes with the availability of future scaling trends.  

\section{Usage of DeepScaleTool and Scaling Factor Computation Examples}
\subsection{DeepScaleTool Usage}
{\begin{figure}
{\centering{\includegraphics[scale = 0.45]{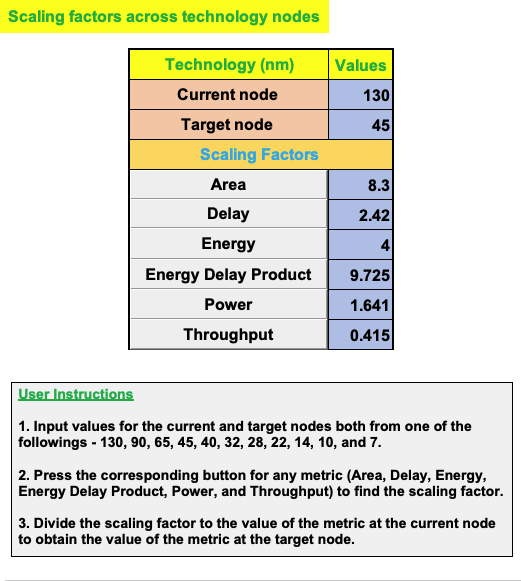}}}
\caption{Screenshot of the DeepScaleTool page showing the values given for current node and target node, generated scaling factors, and user instructions.}
\label{fig:tool}
\end{figure}}
The usage of the DeepScaleTool is simple, which requires three-fold steps as shown in Figure~\ref{fig:tool}. Currently, the tool supports scaling factors for the following fabrication nodes in units of nm: 130, 90, 65, 45, 40, 32, 28, 22, 14, 10, and 7. The user 
inputs 
one of those values for the current node and target node. Next, the user can press the corresponding button for any performance metric or parameter to display the scaling factor. Finally, the value of any metric at a target node can be found 
based on the value of that metric at the current node and resulting scaling factor using the following equation, where $x$ and $y$ denote target node and current node respectively, 
\begin{equation}
Value_x = ~{\ensuremath{Value_y~{/}~Scaling factor}}
\end{equation}
\subsection{Examples of Scaling Factor Computation}
The following examples are shown to illustrate the scaling factor computation procedure. To scale a circuit from 130~nm to 45~nm, area scaling factor is 8.3 per the current version of the tool as shown in Figure~\ref{fig:tool}. If the circuit occupies an area of 100~um\textsuperscript{2} in 130~nm node, the resulting area in 45~nm node using equation~(1) will be 100~/~8.3 = 12.05~um\textsuperscript{2}. Similarly, to scale a circuit from 45~nm to 32~nm, the tool displays a power scaling factor of 1.238. If the circuit dissipates 100~mW in 45~nm node, the resulting power dissipation in 32~nm node using equation~(1) will be 100~/~1.238 = 80.775~mW.

The scaling computations for delay, energy, energy delay product, and throughput can be performed by generating the corresponding scaling factors from the tool and applying equation (1) like the above examples. The scaling factors for derived metrics like throughput{/}area and power density can also be found out using the scaling factors for the corresponding primary metrics that are generated from the tool. 

\section{Comparison of Scaling Factors Estimation Methods and Accuracy with Traditional Scaling}

{\begin{figure}
{\centering{\includegraphics[scale = 0.4]{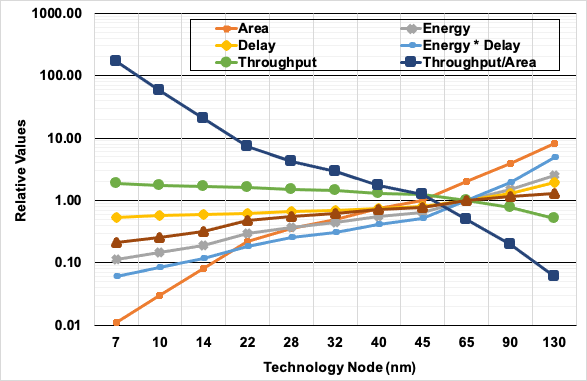}}}
\caption{Scaling trends for area, delay, power from 130~nm to 7~nm based on the modeling presented in this work.}
\label{fig:relative}
\end{figure}}
\subsection{Scaling Trends and Accuracy Analysis with Traditional Scaling}
Figure~\ref{fig:relative} shows the scaling trends for transistor area, delay, energy, throughput, and power based on the modeling presented in this work. Among all the considered parameters transistor area achieves a remarkable scaling over the years, which is better than the traditional scaling trends. The delay and throughput achieves minimal scaling with the recent technology nodes. 
{\begin{figure}
{\centering{\includegraphics[scale = 0.4]{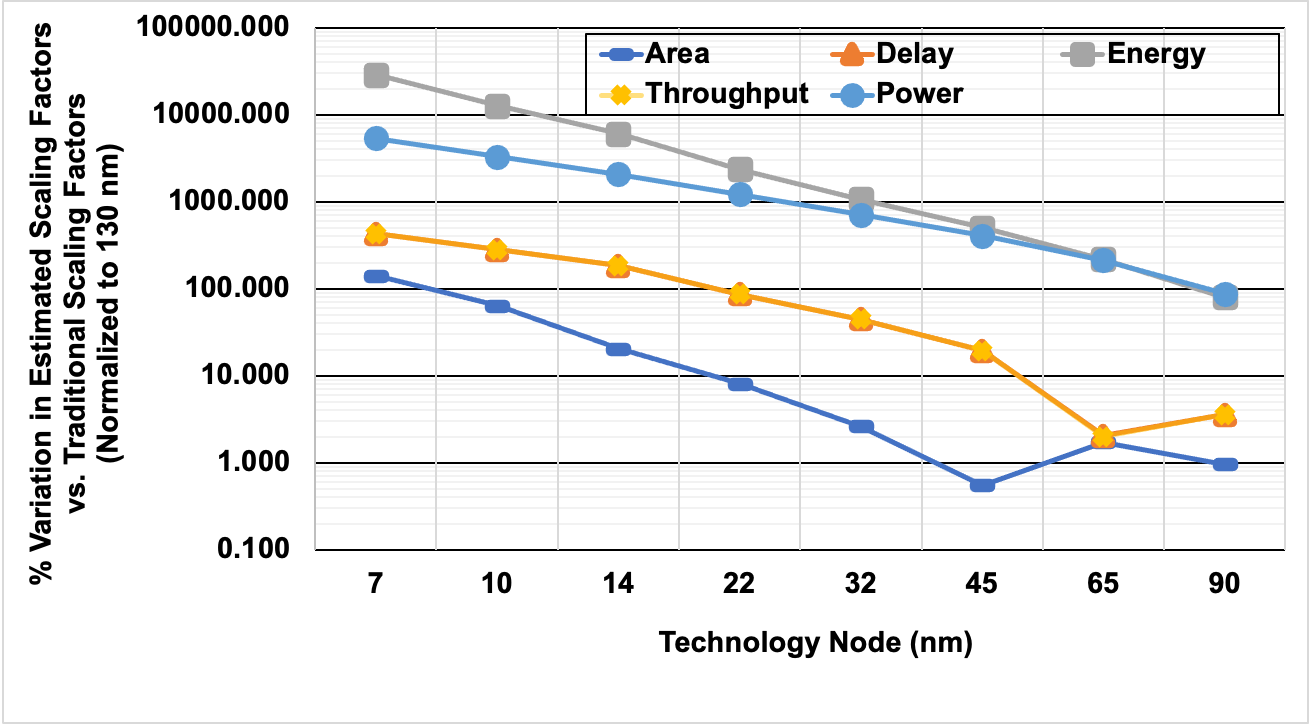}}}
\caption{Traditional scaling factors vs. modeled scaling factors presented in this work for the deep-submicron regime.}
\label{fig:error}
\end{figure}}
Figure~\ref{fig:error} shows the percentage error of value of scaling factors modeled for each parameter with respect to traditional scaling factors. The percentage errors shown for each technology node are relative to the values at 130~nm. Transistor area shows the least variation among all parameters as transistor area still scales by the traditional estimation or even better with the advent of transistor innovations like high-dielectric metal gates, FinFET, and 3D FinFET structures~\cite{Bohr}. However, delay and power dissipation trends vary significantly as compared to traditional scaling improvements of~{\ensuremath{1{/}K}} and~{\ensuremath{1{/}K^2}}. 

Due to the effect of leakage current threshold voltage scaling doesn't happen aggressively and thus the same effect on scaling for supply voltage as well. Therefore, there is a limit on energy efficiency scaling than the traditional scaling factor of~{\ensuremath{1{/}K^3}}. Similarly, transistor gate delay scaling is primarily limited by the slower interconnect scaling. The poor trends in gate delay and power dissipation scaling shown in Figure~\ref{fig:error} affect the energy efficiency scaling and thereby a larger percentage of error is observed for scaling energy. Moreover, as we advance for the recent technology nodes we observe a greater percentage of error in relative scaling values with respect to 130~nm due to the cumulative effect of the error margins across technology nodes. 

\subsection{Comparison of Scaling Factor Estimation Methods}
{\begin{figure}
{\centering{\includegraphics[scale = 0.45]{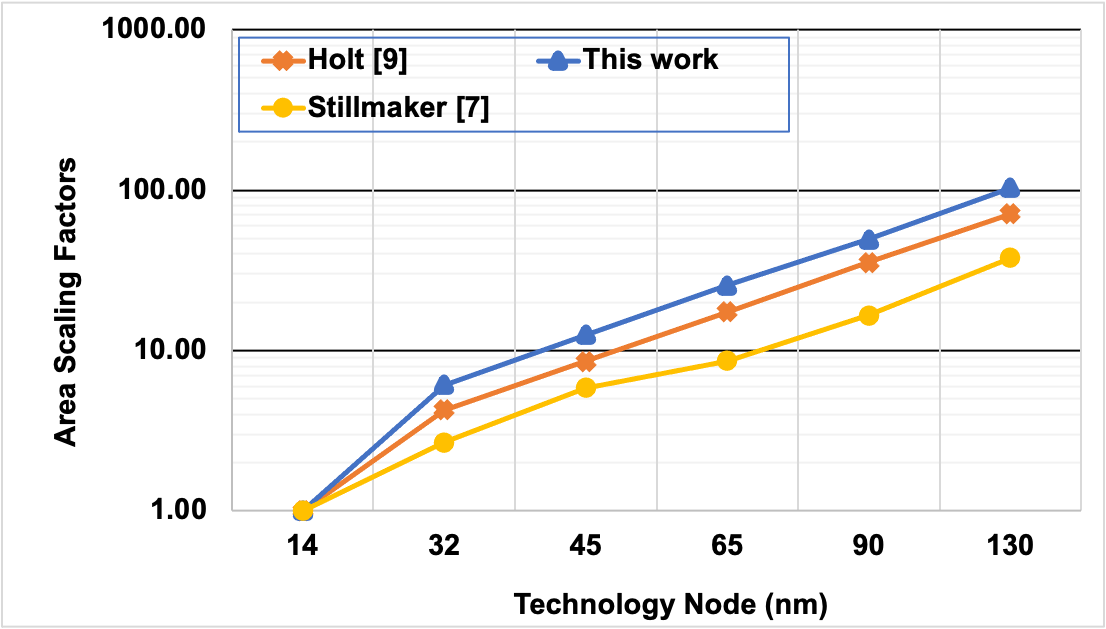}}}
\caption{Comparison of area scaling factors 
between our work and Stillmaker~\cite{Stillmaker201774} for starting nodes from 130~nm to 14~nm with a target node of 14~nm, where in our work shows better correlation to the reference data by Holt~\cite{Holt}.}
\label{fig:areascaling}
\vspace{-1ex}
\end{figure}}

\begin{table}[t]
	\caption{Comparison of area, delay, and power scaling (\% reduction) factors for 10~nm to 7~nm scaling. The reference scaling data~\cite{Cai} corresponds to TSMC's technology scaling.}
	\label{tab:scalingcomparison}
	\vspace{-2ex}
	\begin{center}
		\begin{small}
			\begin{sc}
				\begin{tabular}{l|l|c|c}
					\hline
					 \textbf{Source} & \textbf{Area}&\textbf{Delay} & \textbf{Power}\\

\hline
Cai~\cite{Cai} & 30 -- 35 & 10 & 35\\

\hline
Stillmaker~\cite{Stillmaker201774} \, \, & 59 & 19.1 & 9.1\\
\hline
This work & \textbf{36.7} & \textbf{7.5} & \textbf{30}\\
\hline
\end{tabular}
\end{sc}
\end{small}
\end{center}
\vspace{-1ex}
\end{table}

Table~\ref{tab:scalingcomparison} shows the variation in area, delay, and power percentage scaling from 10~nm to 7~nm. The scaling factor values modeled in this work which are based on Intel's silicon technology scaling trends~\cite{Bohr,Holt} achieve better correlation with TSMC based scaling data~\cite{Cai} than the ITRS and PTM based modeling approach~\cite{Stillmaker201774} in terms of area, delay, and power. The delay and power scaling data for the article~\cite{Stillmaker201774} have been calculated using the given corresponding coefficients, supply voltage per ITRS data, and modeling expressions. The errors in between modeled data presented in this work and TSMC's scaling data are 1\%, 2.5\%, and 5\% for area, delay, and power respectively, which states the reliability of DeepScaleTool across major foundries. However, area, delay, and power scaling factors presented in the article~\cite{Stillmaker201774} differ from TSMC based data by 24--29\%, 9.1\%, and 24.9\% respectively, which is significant given the percentage scaling occurring for those parameters with the said technology nodes. 

Moreover, area scaling factors presented in the article~\cite{Stillmaker201774} varies significantly when compared to modeling based on silicon data as presented by DeepScaleTool as shown in Figure~\ref{fig:areascaling}. For example, 130~nm to 7~nm technology scaling brings down the area by a factor of 110 per the article~\cite{Stillmaker201774}, while the current version of DeepScaleTool suggests the corresponding area scaling factor value of 754.55. The later scaling factor seems more accurate since area scales down by a factor of 
approximately 303 in general over eight generations from 130~nm to 7~nm. Moreover, the aggressive scaling than the normal rate of 0.49 makes the overall factor shoot up to 754.55. Therefore, DeepScaleTool caters to provide more accurate estimation of scaling factors for various design parameters irrespective of major foundries and it avoids the prediction inaccuracy from ITRS and PTM models.  

\section{Conclusion}
We present DeepScaleTool, a tool designed to provide accurate estimation of scaling factors using published silicon trends and polynomial based curve-fitting method. The scaling factors presented in this work shows that the traditional scaling factors go obsolete in deep-submicron era. Although the primary data sets considered for this work belong to Intel's published technology scaling trends, the proposed tool achieves good correlation to the TSMC based scaling trends as well. Moreover, we show that published silicon data based modeling and estimation is more accurate than the simulation based modeling and data per ITRS and PTM, which is the state-of-the-art in estimating scaling factors in the deep-submicron regime. In conclusion, DeepScaleTool provides an easy platform to obtain reliable scaling factors for various design parameters in the deep-submicron era, understand the discrepancies with traditional scaling factors, and also helps in performing fair comparisons of circuits performance over different technology nodes.

\bibliographystyle{IEEEtran} 

\begin{thebibliography}{10}
\providecommand{\url}[1]{#1}
\csname url@samestyle\endcsname
\providecommand{\newblock}{\relax}
\providecommand{\bibinfo}[2]{#2}
\providecommand{\BIBentrySTDinterwordspacing}{\spaceskip=0pt\relax}
\providecommand{\BIBentryALTinterwordstretchfactor}{4}
\providecommand{\BIBentryALTinterwordspacing}{\spaceskip=\fontdimen2\font plus
\BIBentryALTinterwordstretchfactor\fontdimen3\font minus
  \fontdimen4\font\relax}
\providecommand{\BIBforeignlanguage}[2]{{%
\expandafter\ifx\csname l@#1\endcsname\relax
\typeout{** WARNING: IEEEtran.bst: No hyphenation pattern has been}%
\typeout{** loaded for the language `#1'. Using the pattern for}%
\typeout{** the default language instead.}%
\else
\language=\csname l@#1\endcsname
\fi
#2}}
\providecommand{\BIBdecl}{\relax}
\BIBdecl

\bibitem{Moore}
G.~E. {Moore}, ``Cramming more components onto integrated circuits,''
  \emph{Proceedings of the IEEE}, vol.~86, no.~1, pp. 82--85, 1998.

\bibitem{Dennard}
R.~H. {Dennard}, F.~H. {Gaensslen}, H.~{Yu}, V.~L. {Rideout}, E.~{Bassous}, and
  A.~R. {LeBlanc}, ``Design of ion-implanted mosfet's with very small physical
  dimensions,'' \emph{IEEE Journal of Solid-State Circuits}, vol.~9, no.~5, pp.
  256--268, 1974.

\bibitem{Dennard2}
G.~{Baccarani}, M.~R. {Wordeman}, and R.~H. {Dennard}, ``Generalized scaling
  theory and its application to a 1/4 micrometer {MOSFET} design,'' \emph{IEEE
  Transactions on Electron Devices}, vol.~31, no.~4, pp. 452--462, 1984.

\bibitem{rabaey2003digital}
J.~M. Rabaey, A.~P. Chandrakasan, and B.~Nikoli{\'c}, \emph{Digital integrated
  circuits: a design perspective}.\hskip 1em plus 0.5em minus 0.4em\relax
  Pearson Education, 2003, vol.~7.

\bibitem{Uyemura}
J.~Uyemura, \emph{Introduction to VLSI Circuits and Systems}, 1st~ed.\hskip 1em
  plus 0.5em minus 0.4em\relax Hoboken, NJ: John Wiley `|\&' Sons, Inc., 2002.

\bibitem{Bohr}
M.~T. {Bohr} and I.~A. {Young}, ``{CMOS} scaling trends and beyond,''
  \emph{IEEE Micro}, vol.~37, no.~6, pp. 20--29, 2017.

\bibitem{Stillmaker201774}
A.~Stillmaker and B.~Baas, ``Scaling equations for the accurate prediction of
  {CMOS} device performance from 180~nm to 7~nm,'' \emph{Integration, the
  {VLSI} Journal}, vol.~58, pp. 74--81, 2017,
  \url{http://vcl.ece.ucdavis.edu/pubs/2017.02.VLSIintegration.TechScale/}.

\bibitem{Kuhn}
K.~J. {Kuhn}, ``{CMOS} transistor scaling past 32~nm and implications on
  variation,'' in \emph{2010 IEEE/SEMI Advanced Semiconductor Manufacturing
  Conference (ASMC)}, 2010, pp. 241--246.

\bibitem{Holt}
W.~M. {Holt}, ``1.1 moore's law: A path going forward,'' in \emph{2016 IEEE
  International Solid-State Circuits Conference (ISSCC)}, 2016, pp. 8--13.

\bibitem{PTM}
\BIBentryALTinterwordspacing
(2015, Oct.) Predictive technology model. [Online]. Available:
  \url{http://ptm.asu.edu/}
\BIBentrySTDinterwordspacing

\bibitem{ITRS}
\BIBentryALTinterwordspacing
(2015, Oct.) International technology roadmap for semiconductors. [Online].
  Available: \url{http://www.itrs.net/}
\BIBentrySTDinterwordspacing

\bibitem{Cai}
M.~{Cai} \emph{et~al.}, ``7nm mobile soc and 5g platform technology and design
  co-development for ppa and manufacturability,'' in \emph{2019 Symposium on
  VLSI Technology}, 2019, pp. T104--T105.

\bibitem{g3data}
\BIBentryALTinterwordspacing
(2019, Oct.) g3data, a tool for extracting data from scanned graphs. [Online].
  Available: \url{https://github.com/pn2200/g3data}
\BIBentrySTDinterwordspacing

\bibitem{DeepScale}
\BIBentryALTinterwordspacing
S.~{Sarangi} and B.~{Baas}. (2021, Feb.) Deepscaletool. [Online]. Available:
  \url{https://sourceforge.net/projects/deepscaletool/}
\BIBentrySTDinterwordspacing

\end{thebibliography}

\end{document}